\documentclass[11pt]{revtex4}

\newcommand{\N}{N\raise.7ex\hbox{\underline{$\circ $}}$\;$}

\begin{document}

\title{
Some Consequences from the Dirac-K\"{a}hler  Theory:\\
on Intrinsic Spinor  Sub-structure of the  Different Boson Wave
Functions}

\author{E. Ovsiyuk}
\author{O. Veko}
\author{V. Red'kov}

\email{  e.ovsiyuk@mail.ru, vekoolga@mail.ru,
redkov@dragon.bas-net.by}

\affiliation{
 Mozyr State Pedagogical University \\
Institute of Physics, NAS of Belarus}

%\received{}

\begin{abstract}

Properties  of tensors equivalent to  the direct product of two different 4-spinors
are investigated.
It  is shown that  the tensors  obey additional 8 nonlinear restrictions, those are presented in Lorentz
 covariant form.
In the context of the Dirac--K\"{a}hler theory, such a property can be interpreted as follows:
if one wishes to consider
the Dirac--K\"{a}hler  field as consisting of two 4-spinor fields, one must  impose additional restrictions
on tensors of the Dirac--K\"{a}hler  field, which leads to a non-linear wave equation for a complex boson field
(composed  on the base of two 4-spinor fields).

Instead, the use  of four bi-spinor  fields gives possibility to construct the Dirac--K\"{a}hler
tensor set of  16 independent components. However, the formulas relating the Dirac-K\"{a}hler boson to
four fermion fields are completely different from those  previously used in the literature.
In explicit form, restrictions on four 4-spinor corresponding to
separation of different simplest bosons with spin 0 or 1 and various intrinsic parities,
are constructed.

\end{abstract}

\maketitle

\section{Introduction}

The Dirac--K\"{a}hler field  can be described whether as a 2-rank
bispinor or as  a set of tensor fields [1, 2 ].
 Such a
double representation gave rise to many investigations with accent
on equivalence  of the Dirac--K\"{a}hler field
$$
(i \gamma^{a} \partial_{a}- M) U =0 \; , \qquad (i \gamma^{a} \partial_{a}- M) \Psi^{(A)} =0, \qquad  A=1,2,3,4
$$
to four 4-spinor fields
$$
U  = \Psi \otimes \Psi= \left | \begin{array}{cccc}
. & . &. & .\\
. & . &. & .\\. & . &. & .\\. & . &. & .
\end{array} \right | = \{ \Psi^{(1)}, \Psi^{(2)}, \Psi^{(3)}, \Psi^{(4)}\}
$$

We will obtain relationships between four fermion fields and
the Dirac-K\"{a}hler boson, they are  completely different from these (2),  previously used
in the literature (see [1, 2 ] and references therein).

\section{Boson consisted of 2 fermions  }

It is known that the Dirac--K\"{a}hler field can be described
whether as a 2-rank bispinor or as a set of tensor
fields 9see notation in [2])
$$
U  = \Psi \otimes \Psi  = \left (  - i  \Phi  + \gamma^{b} \Phi
_{b}  +
           i  \sigma^{ab}  \Phi _{ab}  +  \gamma ^{5}  \tilde{\Phi }  +
           i  \gamma ^{b} \gamma ^{5}  \tilde {\Phi }_{b}   \right ) E^{-1} \; ;
\eqno(1a)
$$

\noindent pseudo-quantities are referred by a tilde symbol. Such a
double representation gave rise to many investigations  (if not
speculations) with accent on equivalence of the Dirac--K\"{a}hler
field to four bispinor fields. In this connection we will consider
the following problem: let it will given a direct product of two
arbitrary and different 4-spinor fields (thereby we have a system
consisting of two  fermions)
$$
\Phi  = \left | \begin{array}{c}
\xi^{1} \\ \xi^{2} \\
\eta_{\dot{1}} \\  \eta_{\dot{2}}
\end{array} \right |=
\left | \begin{array}{c}
A \\ B \\
C \\  D
\end{array} \right |, \qquad \Psi  =
\left | \begin{array}{c}
\Sigma^{1} \\ \Sigma^{2} \\
H_{\dot{1}} \\  H_{\dot{2}}
\end{array} \right |=
\left | \begin{array}{c}
M \\ N \\
K \\  L
\end{array} \right |,
$$
$$
 U   = \Phi \otimes \Psi =
 \left | \begin{array}{rrrr}
  A M  &  A N  &   A K  &   A L \\
  B M  &  B N  &   B K  &   B L \\
  C M  &  C N  &   C K  &   C L \\
  D M  &  D N  &   D K  &   D L  \\
\end{array} \right |  = \left | \begin{array}{cc}
\xi & \Delta \\
W & \eta
\end{array} \right | .
\eqno(1b)
$$

\noindent In 2-spinor form, expansion (1a)  looks as follows
$$
\Delta (x) = [ \;\Psi _{l}(x) + i \; \tilde{\Psi}_{l}(x) \; ]\;
\bar{\sigma} ^{l} \dot{\epsilon } \; , \qquad W(x) = [ \;\Psi
_{l}(x) - i \; \tilde{\Psi }_{l}(x))\;] \; \bar{\sigma}^{l}
\epsilon ^{-1} \; ,
$$
$$
\xi (x) = [ \; - i \;  \Psi (x) - \tilde{\Psi} (x) + i \; \Sigma
^{mn} \; \Psi _{mn}(x)\;  ] \; \epsilon ^{-1} \; ,\qquad \eta (x)
= [\; - i \; \Psi (x) + \tilde{\Psi} (x) +
 i\;  \bar{\Sigma}^{mn}\; \Psi _{mn}(x)\; ]\; \dot{\epsilon } \;  .
\eqno(2)
$$

\noindent and inverse relations are
$$
\Psi ^{l}(x) + i \; \tilde{\Psi} ^{l}(x)  = {1\over 2} \;
\mbox{Sp}\;  [\; \dot{\epsilon }^{-1} \; \sigma ^{l} \;\Delta(x)
\; ] \; ,\qquad \Psi ^{l}(x) - i \; \tilde{\Psi }^{l}(x)  =
 {1\over 2} \; \mbox{Sp} \;  [\;  \epsilon \bar{\sigma}^{l} \; W(x)\; ] \; ,
\eqno(3a)
$$
$$
-i \; \Psi(x)  - \tilde{\Psi}(x)   = {1\over 2} \; \mbox{Sp}\; [\;
\epsilon  \; \xi(x)\;  ] \; ,\qquad
 -i \; \Psi(x)  + \tilde{\Psi
}(x)  = {1\over 2} \; \mbox{Sp} \; [ \; \dot{\epsilon }^{-1} \;
\eta(x) \; ] \; , \eqno(3b)
$$
$$
  - i \; \Psi ^{kl}(x) + {1\over 2}\; \epsilon ^{klmn} \;\Psi _{mn}(x) =
\mbox{Sp}\; [ \; \epsilon \;  \Sigma ^{kl} \xi(x)\; ] \; ,
$$
$$
 - i\;  \Psi ^{kl}(x) - {1\over 2}\; \epsilon ^{klmn}\; \Psi _{mn}(x) =
\mbox{Sp} \; [ \; \dot{\epsilon }^{-1} \; \bar{\Sigma}^{kl} \;
\eta(x) \; ] \; . \eqno(3c)
$$

\noindent Remind the notation:
$$
\sigma^{a} = (I, + \sigma^{j}) \; , \qquad \sigma_{a} = (I, -
\sigma^{j}) \; ,
$$
$$
\epsilon =  +i \;  \sigma^{2} = \left | \begin{array}{cc} 0& +1
\\  -1 & 0 \end{array} \right |,
\qquad \epsilon^{-1} =  - i \sigma^{2}= \left | \begin{array}{cc}
 0& -1 \\ +1 & 0 \end{array} \right |.
$$
$$
\dot{\epsilon}  = + i \sigma^{2}= \left | \begin{array}{cc} 0& +1
\\  -1 & 0 \end{array} \right |\;, \qquad
(\dot{\epsilon} )^{-1} =  - i\sigma^{2}=\left | \begin{array}{cc}
 0& -1 \\ +1 & 0 \end{array} \right |\; .
$$

 Specifying the formulas  (3c)--(3c), we get
 $$
 \Psi =-  {1\over 4i} \; (BM-AN  + CL - DK) \; , \qquad
 \tilde{\Psi}  =  - {1\over 4}\;   (BM-AN  - CL + DK) \; .
\eqno(4)
$$

\noindent
 Vector and pseudo-vector are
$$
\Phi^{0} = {1 \over 4}\;   (AL-BK  + DM-CN)\;   ,\qquad
\tilde{\Phi}^{0} = {1 \over 4i} \; (AL-BK - DM + CN)\;  ,
$$
$$
\Psi^{1} = -{1 \over 4} \; (AK-BL  + CM-DN )\; , \qquad
\tilde{\Psi}^{1} =- {1 \over 4i}  \; (AK-BL   - CM + DN)\; ,
$$
$$
\Phi^{2} = -{i \over 4} \;  (AK+BL + CM+DN )\;  , \qquad
\tilde{\Phi}^{2} =- {1 \over 4} \;  (AK+BL - CM- DN)\;  ,
$$
$$
\Phi^{3} = {1 \over 4}\;  (BK+AL + DM+CN)\; , \qquad
\tilde{\Phi}^{3} = {1 \over 4i}\; (BK+AL  - DM- CN)\; .
\eqno(5)
$$

\noindent An identity folds
$$
\Psi^{a} \tilde{\Psi}_{a} =
 \Psi^{0} \tilde{\Psi}^{0}-
 \Psi^{1} \tilde{\Psi}^{1} - \Psi^{2}\tilde{\Psi}^{2} - \Psi^{3}  \tilde{\Psi}^{3}
=0\;.  \eqno(6)
$$

    Now, let us turn to antisymmetric tensor:
 $$
   \Psi ^{01} +  i \Psi ^{23} = -{i \over 2}(BN-AM)
   \; , \qquad
  \Psi ^{01} -  i  \Psi ^{23} =- {i \over 2}(D L -CK)
  \; ,
$$
$$
   \Psi ^{02} +  i \Psi ^{31} = -{1 \over 2}(AM+BN)\; ,
\qquad
   \Psi ^{02} -  i \Psi^{31} = -{1 \over 2}(CK+DL) \; ,
$$
$$
  \Psi ^{03} +  i  \Psi ^{12} = -{i \over 2}(AN+BM)\; ,
\qquad
  \Psi ^{03} - i  \Psi ^{12} =- {i \over 2}( CL+DK)  \; ;
\eqno(7)$$

\noindent which lead to
$$
\Psi ^{01} ={i \over 4}(AM-BN+CK-DL)\; , \qquad \Psi ^{23}={1
\over 4}(AM-BN-CK+DL) \; ,
$$
$$
\Psi ^{02} =-{1 \over 4}(AM+BN+CK+DL)\; ,\qquad \Psi ^{31}={i
\over 4}(AM+BN-CK-DL) \; ,
$$
$$
\Psi ^{03} =-{i \over 4}(AN+BM+CL+DK)\; ,\qquad  \Psi ^{12}=-{1
\over 4}(AN+BM-CL-DK) \; . \eqno(8)
$$

Let us consider  possibilities for vanishing some of these tensor
constituents relevant to  $\Phi \otimes \Psi$. First, let scalar
and pseudo-scalar be zero
$$
 \Psi =-  {1\over 4i} \; (BM-AN  + CL - DK)  = 0\; ,\qquad
 \tilde{\Psi}  =  - {1\over 4}\;   (BM-AN  - CL + DK)  = 0 \; ;
\eqno(9a)
$$

\noindent these equation are equivalent to more simple ones
$$
BM-AN=0 \qquad\mbox{or}  \qquad \xi^{1} \Sigma^{2} - \xi^{2}
\Sigma^{1} = 0 \qquad\mbox{or}  \qquad {\xi^{1} \over \Sigma^{1}}
=  {\xi^{2} \over  \Sigma^{2} }\; , \;\;\;
$$
$$
CL - DK=0 \qquad \mbox{or}\qquad  \eta_{\dot{1}}  H_{\dot{2}} -
\eta_{\dot{2}}  H_{\dot{1}}=0 \qquad\mbox{or}  \qquad {
\eta_{\dot{1}} \over H_{\dot{1}} } = { \eta_{\dot{2}}\over
H_{\dot{2}}} \; , \eqno(9b)
$$

\noindent It means that 2-spinors must be  proportional to each other
$$
\Sigma^{\alpha} (x)  = \mu \; \xi (x) \;, \qquad
H_{\dot{\alpha}}(x)  = \nu  \; \eta_{\dot{\alpha}}(x) \; ;
\eqno(10a)
$$

\noindent Linear restrictions  (10a) can be presented as
$$
M = \mu \; A \;,\qquad  N = \mu \; B \; , \qquad K = \nu \; C \;,
\qquad L = \nu \; D \; . \eqno(10b)
$$

\noindent At this
$$
\Psi (x) = 0\; , \qquad \tilde{\Psi} (x) = 0 \; , \eqno(11a)
$$

\noindent and
$$
\Phi^{0} = {1 \over 4}\; (AD -BC) (\nu + \mu) \;   , \qquad
\tilde{\Phi}^{0} = {1 \over 4i} (AD -BC) (\nu - \mu)\; \;  ,
$$
$$
\Psi^{1} = -{1 \over 4} (AC -BD)(\nu + \mu)  \; , \qquad
\tilde{\Psi}^{1} = - {1 \over 4i}  (AC-BD)( \nu - \mu ) \; ,
$$
$$
\Phi^{2} =  -{i \over 4} \;  (AC+BD) (\nu + \mu) \;  , \qquad
\tilde{\Phi}^{2} = - {1 \over 4} \; (AC+BD) (\nu - \mu) \;  ,
$$
$$
\Phi^{3} = + {1 \over 4}\; (BC+AD) (\nu + \mu) \; ,\qquad
\tilde{\Phi}^{3} = +{1 \over 4i}\;(BC+AD) (\nu - \mu)\; .
$$

\noindent Note that vector and pseudovector are proportional to each other
$$
(\nu - \mu )  \Phi^{a} = i \;(\nu + \mu) \tilde{\Phi}^{a} \; .
\eqno(11b)
$$
In turn,   3-vectors related to antisymmetric tensor take the form
$$
 s_{1} =   \Psi ^{01} +  i \Psi ^{23} = -{i \over 2}(B^{2} -A^{2}) \; \mu
   \; , \qquad
 t_{1} =  \Psi ^{01} -  i  \Psi ^{23} =- {i \over 2}(D^{2} -C^{2}) \; \nu
  \; .
$$
$$
 s_{2}=   \Psi ^{02} +  i \Psi ^{31} = -{1 \over 2}(A^{2} +B^{2})\;  \mu \; ,
\qquad
 t_{2}=   \Psi ^{02} -  i \Psi^{31} = -{1 \over 2}(C^{2} +D^{2}) \; \nu  \; .
$$
$$
 s_{3}=  \Psi ^{03} +  i  \Psi ^{12} = -i \; A  B \; \mu \; ,
\qquad
 t_{3} =  \Psi ^{03} - i  \Psi ^{12} =- i \; C D \;  \nu   \; ;
\eqno(11c)
$$

 \noindent
they are isotropic and non-orthogonal to each other
$$
{\bf s}^{2} = 0\;, \qquad  {\bf t}^{2} = 0\;, \qquad {\bf s}\;
{\bf t} = {\mu \nu \over 2} (AD -BC)^{2} \; . \eqno(11d)
$$

On may impose more weak  requirements. For instance, let it be
\underline{$\Psi \neq 0 , \tilde{\Psi} =0 $}, then
$$
BM-AN  + CL - DK \neq 0\; , \qquad
 BM-AN  - CL + DK = 0\; ,
$$
so that
$$
 BM-AN  = +(CL - DK)  \neq  0\; , \qquad
\xi^{2}\Sigma^{1}-\xi^{1} \Sigma^{2}  =  +(\eta_{\dot{1}}
H_{\dot{2}} -  \eta_{\dot{2}} H_{\dot{1}}) \neq  0 \; .
\eqno(12)
$$

\noindent Let it be  \underline{$\Psi = 0 , \tilde{\Psi} \neq 0 $},
then
$$
BM-AN  + CL - DK = 0\; , \qquad
 BM-AN  - CL + DK \neq  0\; ,
$$
that is
$$
BM-AN  = - ( CL - DK)  \; ,\qquad  \xi^{2} \Sigma^{1}
-\xi^{1}\Sigma^{2}  = - (\eta_{\dot{1}}H_{\dot{2}} -
\eta_{\dot{2}}H_{\dot{1}})  \neq  0 \; .
 \eqno(13)
$$

Now, let us consider the case of vanishing two 4-vectors
(see (5b)):
$$
\Psi^{a}=0\;, \qquad  \tilde{\Psi}^{a}=0 \; ;
$$
$$
AL-BK  + DM-CN=0\;   ,\qquad AL-BK - DM + CN=0\;  ,
$$
$$
AK-BL  + CM-DN =0\; , \qquad AK-BL   - CM + DN = 0\; ,
$$
$$
AK+BL + CM+DN =0 \;  , \qquad AK+BL - CM- DN =0 \;  ,
$$
$$
BK+AL + DM+CN =0 \; , \qquad BK+AL  - DM- CN = 0\; ;
 \eqno(14a)
 $$

\noindent which is equivalent to

$$
AL-BK  =0 \; , \qquad DM-CN=0\;   , \qquad AK-BL =0 \; , \qquad
CM-DN =0\; ,
$$
$$
AK+BL=0 \; , \qquad  CM+DN =0 \;  , \qquad  BK+AL =0 \; , \qquad
DM+CN =0 \; ,
$$

\noindent these lead to
$$
AK=0\;, \qquad BL = 0\; , \qquad DM= 0 \; , \qquad CN=0 \; ,
$$
$$
AL=0 \;, \qquad BK = 0 \; , \qquad CM =0 \; , \qquad DN =0\; .
 \eqno(14b)
$$
There exist two solutions (however, they are Lorentz invariant
only with respect to continuous transformations)
$$
1) \qquad  A=0\;,\; B=0\;, \;M=0 \; , \;N=0\; , \qquad
\xi^{\alpha}=0\;, \; \Sigma^{\alpha}=0 \; ;
$$
$$
2) \qquad  C=0\;,\; D=0\;, \;K=0 \; , \;L=0\; , \qquad
\eta_{\dot{\alpha}}=0 \;,  \; H_{\dot{\alpha}}=0 )\; .
 \eqno(15a)
$$

\noindent The lead respectively to
\vspace{2mm}

the case  1)
$$
 \Psi =  {i\over 4} \; ( CL - DK) \; ,\qquad
 \tilde{\Psi}  =   {1\over 4}\;   (  CL - DK) \; ,
$$
$$
   \Psi ^{01} +  i \Psi ^{23} = 0
   \; , \qquad
  \Psi ^{01} -  i  \Psi ^{23} =- {i \over 2}(D L -CK)
  \; ,
$$
$$
   \Psi ^{02} +  i \Psi ^{31} =0\; ,
\qquad
   \Psi ^{02} -  i \Psi^{31} = -{1 \over 2}(CK+DL) \; ,
$$
$$
  \Psi ^{03} +  i  \Psi ^{12} = 0\; ,
\qquad
  \Psi ^{03} - i  \Psi ^{12} =- {i \over 2}( CL+DK)  \; ;
 \eqno(15b)
$$

the case  2)
$$
 \Psi =  {i\over 4} \; (BM-AN  ) \; , \qquad
 \tilde{\Psi}  =  - {1\over 4}\;   (BM-AN  ) \; ,
$$
$$
   \Psi ^{01} +  i \Psi ^{23} = -{i \over 2}(BN-AM)
   \; , \qquad
  \Psi ^{01} -  i  \Psi ^{23} = 0
  \; ,
$$
$$
   \Psi ^{02} +  i \Psi ^{31} = -{1 \over 2}(AM+BN)\; ,
\qquad
   \Psi ^{02} -  i \Psi^{31} = 0 \; ,
$$
$$
  \Psi ^{03} +  i  \Psi ^{12} = -{i \over 2}(AN+BM)\; ,
\qquad
  \Psi ^{03} - i  \Psi ^{12} = 0   \; .
 \eqno(15c)
$$

There is possible to impose more weak restriction. For instance,
let it be $\tilde{\Psi}^{a}=0$:
$$
\Phi^{0} = {1 \over 2}\;   (AL-BK )\;   ,\qquad AL-BK  =  +(DM -
CN )\;  ,
$$
$$
\Psi^{1} = -{1 \over 2} \; (AK-BL )\; , \qquad AK-BL =  +(CM - DN)
\; ,
$$
$$
\Phi^{2} = -{i \over 2} \;  (AK+BL )\;  , \qquad AK+BL  =  +(CM +
DN )\;  ,
$$
$$
\Phi^{3} = {1 \over 2}\;  (BK+AL )\; , \qquad BK+AL =  +(DM +  CN)
\; ; \eqno(16a)
$$

\noindent which is equivalent to
$$
AK = CM\; , \qquad BL = DN\;, \qquad  AL = DM\;, \qquad BK = CN \;
$$

\noindent or
$$
{A \over M} = {C \over K} \qquad \Longrightarrow \qquad {\xi^{1}
\over \Sigma^{1}} = { \eta_{\dot{1}} \over H _{\dot{1}}}\; ,\qquad
{B \over N} = {D \over L} \qquad \Longrightarrow \qquad {\xi^{2}
\over \Sigma^{2}} = { \eta_{\dot{2}} \over H _{\dot{2}}}\; ,
$$
$$
{A \over M} = {D \over L} \qquad \Longrightarrow \qquad {\xi^{1}
\over \Sigma^{1}} = { \eta_{\dot{2}} \over H _{\dot{2}}}\;, \qquad
{B \over N} = {C \over K} \qquad \Longrightarrow \qquad {\xi^{2}
\over \Sigma^{2}} = { \eta_{\dot{1}} \over H _{\dot{1}}}
$$

\noindent or even shorter
$$
{\xi^{1} \over \Sigma^{1}} = {\xi^{2} \over \Sigma^{2}} =  {
\eta_{\dot{1}} \over H _{\dot{1}}} = { \eta_{\dot{2}} \over H
_{\dot{2}}} \qquad \Longleftrightarrow \qquad \Sigma ^{\alpha} =
\mu \xi^{\alpha} \; ,  \qquad  H_{\dot{\alpha}}  = \mu
\eta_{\dot{\alpha}}\; . \eqno(16b)
$$

\noindent It means that two 4-spinors are proportional to each other.

Now, let us impose another restriction, $\Psi^{a}=0$  (see
(5b)):
$$
AL-BK = - (DM-CN )\;   ,\qquad \tilde{\Phi}^{0} = {1 \over 2i} \;
(AL-BK )\;  ,
$$
$$
AK-BL = - (CM-DN) \; , \qquad \tilde{\Psi}^{1} =- {1 \over 2i}  \;
(AK-BL  )\; ,
$$
$$
AK+BL = - ( CM+DN )\;   , \qquad \tilde{\Phi}^{2} =- {1 \over 2}
\;  (AK+BL)\;  ,
$$
$$
BK+AL = - ( DM+ CN)\; , \qquad \tilde{\Phi}^{3} = {1 \over 2i}\;
(BK+AL )\; , \eqno(17a)
$$

\noindent which is equivalent to
$$
AL = - DM \; , \qquad BK = - CN\;, \qquad  AK = - CM\;, \qquad  BL
= - DN \; ;
$$

\noindent or
$$
{A \over M} = -{C \over K} \qquad \Longrightarrow \qquad {\xi^{1}
\over \Sigma^{1}} =  -{ \eta_{\dot{1}} \over H _{\dot{1}}}\;,
\qquad {B \over N} = -{D \over L} \qquad \Longrightarrow \qquad
{\xi^{2} \over \Sigma^{2}} = - { \eta_{\dot{2}} \over H
_{\dot{2}}}
$$
$$
{A \over M} = -{D \over L} \qquad \Longrightarrow \qquad {\xi^{1}
\over \Sigma^{1}} = - { \eta_{\dot{2}} \over H _{\dot{2}}}\; ,
\qquad {B \over N} = -{C \over K} \qquad \Longrightarrow \qquad
{\xi^{2} \over \Sigma^{2}} = -  { \eta_{\dot{1}} \over H
_{\dot{1}}}\; ,
$$

\noindent and shorter
$$
{\xi^{1} \over \Sigma^{1}} = {\xi^{2} \over \Sigma^{2}} =  -{
\eta_{\dot{1}} \over H _{\dot{1}}} = -{ \eta_{\dot{2}} \over H
_{\dot{2}}} \qquad \Longleftrightarrow \qquad \Sigma ^{\alpha} =
\mu \xi^{\alpha} \; ,  \qquad  H_{\dot{\alpha}}  = -\; \mu
\eta_{\dot{\alpha}}\; . \eqno(17b)
$$

Equations (17b)  describe system invariant only with respect to continuous transformations.
Relations  (17b may be written in more short form
$$
{M \over A} = \mu\; , \qquad {N\over B} = \mu \; , \qquad {K\over
C } = -\mu \; ,\qquad
 {L\over D} = - \mu \; .
 $$

 \noindent Corresponding  scalar and pseudo-scalars are
 $$
 \Psi =0 \; , \qquad
 \tilde{\Psi}  = 0 \; .
\eqno(17c)
$$
$$
   \Psi ^{01} +  i \Psi ^{23} = -{i \over 2} \mu (BB-AA) \; ,\qquad
  \Psi ^{01} -  i  \Psi ^{23} =+ {i \over 2} \mu (DD-CC)\; ,
$$
$$
   \Psi ^{02} +  i \Psi ^{31} =-{ 1\over 2}\mu (BB+AA)\; , \qquad
   \Psi ^{02} -  i \Psi^{31} = + {i \over 2} \mu (DD+CC) \; ,
$$
$$
  \Psi ^{03} +  i  \Psi ^{12} = -iAN=-iBM\; , \qquad
  \Psi ^{03} - i  \Psi ^{12} =-i CL=-iDK  \; .
\eqno(17d)
$$

Finally, let us consider the case of vanishing tensor $\Phi^{ab} =
0$:
$$
BN-AM =0   \; , \qquad
  D L -CK= 0
  \; .
$$
$$
 AM+BN =0 \; ,
\qquad
 CK+DL= 0 \; ,
$$
$$
AN+BM = 0\; , \qquad
  CL+DK =0   \; ;
\eqno(18a)
$$

\noindent these conditions are equivalent to
$$
BN=0 \; , \qquad AM  =0   \; , \qquad
  D L =0\; , \qquad CK = 0
  \; ,
$$
$$
BM =-AN\; , \qquad
  CL=- DK   \; ;
\eqno(18b)
$$

There exist two different solutions
-- they both are Lorentz invariant only with respect to continuous transformations
$$
1) \qquad A=0\; , \; B = 0 \;, \; K=0 \; , \; L=0 \; , \qquad
\qquad  \xi^{\alpha} =0 \; , \; H_{\dot{\alpha}} = 0\; ,
$$
$$
1) \qquad C=0\; , \; D = 0\; , \; M=0 \; , \; N=0 \; ,  \qquad
\qquad \eta_{\dot{\alpha}} =0 \; , \; \Sigma^{\alpha} = 0 \; .
\eqno(18c)
$$

\noindent At this, remaining constituents become much simpler (see  (4b)  and  (5b)):

1)
$$
 \Psi =0 \; , \qquad
 \tilde{\Psi}  = 0 \; ,\qquad \tilde{\Psi} ^{a} = + i \Psi \; ,
$$
$$
\Phi^{0} = {1 \over 4}\;   ( DM-CN)\;   ,\qquad \tilde{\Phi}^{0} =
{i \over 4} \; ( DM - CN)\;  ,
$$
$$
\Psi^{1} = {1 \over 4} \; (DN- CM )\; , \qquad \tilde{\Psi}^{1} =
{i  \over 4}  \; (   DN - CM )\; ,
$$
$$
\Phi^{2} = -{i \over 4} \;  ( CM+DN )\;  , \qquad \tilde{\Phi}^{2}
=  {1 \over 4} \;  ( CM+ DN)\;  ,
$$
$$
\Phi^{3} = {1 \over 4}\;  ( DM+CN)\; , \qquad \tilde{\Phi}^{3} =
{i \over 4}\; (  DM + CN)\; ; \eqno(18d)
$$

2)
$$
 \Psi = 0 \;, \qquad
 \tilde{\Psi}  =  0\; , \qquad \tilde{\Psi} ^{a} = - i \Psi \; ,
 $$
$$
\Phi^{0} = {1 \over 4}\;   (AL-BK )\;   ,\qquad \tilde{\Phi}^{0} =
- {i \over 4} \; (AL-BK)\;  ,
$$
$$
\Psi^{1} = -{1 \over 4} \; (AK-BL )\; , \qquad \tilde{\Psi}^{1} =
{i \over 4}  \; (AK-BL  )\; ,
$$
$$
\Phi^{2} = -{i \over 4} \;  (AK+BL  )\;  , \qquad \tilde{\Phi}^{2}
=- {1 \over 4} \;  (AK+BL )\;  ,
$$
$$
\Phi^{3} = {1 \over 4}\;  (BK+AL )\; , \qquad \tilde{\Phi}^{3} =
-{i \over 4}\; (BK+AL )\; . \eqno(18e)
$$

Because according to the process consider above from 8 independent
complex  quantities, constituents of two 4-spinors, there are
constructed 16  complex-valued tensor components, we should expect
existence of additional relations to which these tensors must
obey. By direct calculation, we verify identities
 $$
 \Psi^{ab} \Psi_{b}=  -  \;\tilde{\Psi} \; \tilde{\Psi^{a}}\; ,
\qquad
 \Psi^{ab} \tilde{\Psi}_{b} =+ \tilde{\Psi} \Psi^{a} \; ;
\eqno(19a)
$$

\noindent  besides, note identity
$$
\Psi^{a} \tilde{\Psi}_{a} = 0\; . \eqno(19b)
$$

In connection with  (19)  we can speculate about physical meaning
of the Dirac-K\"{a}hler theory. Indeed, in tensor form this system
is governed by the linear system
 $$
\partial _{l}\Psi  + m \Psi _{l} = 0 \; , \qquad
\partial _{l} \tilde{\Psi } + m \tilde{\Psi }_{l} = 0 \; ,\qquad
\partial _{l} \Psi  + \partial_{a} \Psi _{l}^{\;\;a} - m \Psi _{l} = 0 \; ,
$$
$$
\partial _{l} \tilde{\Psi } -
{1\over 2} \;\epsilon ^{\;\;amn}_{l} \; \partial _{a} \Psi _{mn} -
m \tilde{\Psi }_{l}  = 0  \; ,\qquad
\partial _{m} \Psi _{n} - \partial _{n} \Psi _{m} +
\epsilon ^{\;\;\;\;ab}_{mn} \;\partial _{a} \tilde{\Psi } _{b} - m
\Psi _{mn} = 0 \; . \eqno(20a)
$$

In assumption that the Dirac-K\"{a}hler tensors are constructed
in terms of two 4-spinor according to (1),
we  must assume constraints (19) on these tensors
 -- at this the system  (20a) results in
$$
\partial _{l}\Psi  +  m { \Psi_{ln} \over +\tilde{ \Psi}} \tilde{\Psi}^{n}= 0 \; , \qquad
 \partial _{l} \tilde{\Psi } +  m    {\Psi_{ln}\over -\tilde{ \Psi}} \Psi^{n} = 0 \; ,
 \qquad
  \partial _{l} \Psi   -\tilde{\Psi}  \partial_{a} \Psi _{l}^{\;\;a} -  m
   {\Psi_{ln}  \over +\tilde{ \Psi}} \tilde{\Psi}^{n} = 0 \; ,
$$
$$
 \partial _{l} \tilde{\Psi } -
{1\over 2} \;\epsilon ^{\;\;amn}_{l} \; \partial _{a} \Psi _{mn}
-  m {  \Psi_{ln} \over -\tilde{ \Psi}} \Psi^{n}  = 0  \; ,\qquad
\partial _{m} \Psi _{n} - \partial _{n} \Psi _{m} +
\epsilon ^{\;\;\;\;ab}_{mn} \;\partial _{a} \tilde{\Psi } _{b} - m
\Psi _{mn} = 0 \; .
\eqno(20b)
$$

\noindent
In other word, here we have non-linear boson equations referring  to a couple of starting fermion fields
It may be specially stressed that additional constrains are lorentz invariant.

\section{Boson consisted of 4 fermions}

Let us examine a 2-rank bispinor, consisting offour different 4-spinors
according to the rule
$$
U'' = \Phi \otimes \Psi + \Phi' \otimes \Psi'
$$

\noindent possible coefficients in this combination can be
eliminated by relevant redifinitions. Such an object from the very
beginning contains 16 independent components, so one can expect 16
independent tensor components of the Dirac-K\"{a}hler field.  Let
us use the notation
$$
\Phi  =
\left | \begin{array}{c}
A \\ B \\
C \\  D
\end{array} \right |, \qquad \Psi  =
\left | \begin{array}{c}
M \\ N \\
K \\  L
\end{array} \right |, \qquad
\Phi'  =
\left | \begin{array}{c}
A' \\ B' \\
C' \\  D'
\end{array} \right |, \qquad \Psi'  =
\left | \begin{array}{c}
M' \\ N '\\
K' \\  L'
\end{array} \right |,
$$
$$
U   = \Phi \otimes \Psi =
 \left | \begin{array}{rrrr}
  A M  &  A N  &   A K  &   A L \\
  B M  &  B N  &   B K  &   B L \\
  C M  &  C N  &   C K  &   C L \\
  D M  &  D N  &   D K  &   D L  \\
\end{array} \right |  = \left | \begin{array}{cc}
\xi & \Delta \\
W & \eta
\end{array} \right |,
$$
$$
U '  = \Phi' \otimes \Psi' =
 \left | \begin{array}{rrrr}
  A' M'  &  A' N'  &   A' K'  &   A' L '\\
  B' M'  &  B' N'  &   B' K'  &   B' L' \\
  C' M'  &  C' N'  &   C' K''  &  C' L' \\
  D' M'  &  D' N'  &   D' K'  &   D' L'  \\
\end{array} \right |  = \left | \begin{array}{cc}
\xi' & \Delta' \\
W' & \eta'
\end{array} \right |.
\eqno(21)
$$

\noindent Now, instead of (3) we have
$$
\Psi ^{l}(x) + i \; \tilde{\Psi} ^{l}(x)  = {1\over 2} \; \mbox{Sp}\;  [\;
\dot{\epsilon }^{-1} \; \sigma ^{l} \;( \Delta + \Delta ')  \; ] \; ,\;\;\;
\Psi ^{l}(x) - i \; \tilde{\Psi }^{l}(x)  =
 {1\over 2} \; \mbox{Sp} \;  [\;  \epsilon \bar{\sigma}^{l} \; (W +W')\; ] \; ,
\eqno(22a)
$$
$$
-i \; \Psi(x)  - \tilde{\Psi}(x)   = {1\over 2} \; \mbox{Sp}\;  [\;
\epsilon  \; (\xi + \xi')\;  ] \; ,\;\;\;
 -i \; \Psi(x)  + \tilde{\Psi
}(x)  = {1\over 2} \; \mbox{Sp} \; [ \; \dot{\epsilon }^{-1} \; (\eta + \eta ') \;
] \; ,
\eqno(22b)
$$
$$
  - i \; \Psi ^{kl}(x) + {1\over 2}\; \epsilon ^{klmn} \;\Psi _{mn}(x) =
\mbox{Sp}\; [ \; \epsilon \;  \Sigma ^{kl} ( \xi + \xi') \; ] \; ,
$$
$$
 - i\;  \Psi ^{kl}(x) - {1\over 2}\; \epsilon ^{klmn}\; \Psi _{mn}(x) =
\mbox{Sp} \; [ \; \dot{\epsilon }^{-1} \; \bar{\Sigma}^{kl} \;  (\eta + \eta ') \;
] \; .
\eqno(22c)
$$

Explicitly, equivalent  tensor constituents look
$$
 \Psi'' =-  {1\over 4i} \; [ \; (BM-AN  + CL - DK) + (B'M'-A'N'  + C'L' - D'K')\; ] = \Psi + \Psi '\;
 ,\qquad
 $$
$$
 \tilde{\Psi}''  =  - {1\over 4}\;  [\; (BM-AN  - CL + DK) +  (B'M'-A'N'  - C'L' + D'K')\; ]= \tilde{\Psi} + \tilde{\Psi}' \; .
\eqno(23a)
$$

\noindent 4-vectors
$$
\Phi^{''0} = {1 \over 4}\;   [\; (AL-BK  + DM-CN) + [\; (A'L'-B'K'  + D'M'-C'N') \; ]= \Phi^{0}+\Phi^{'0} \;   ,
$$
$$
\tilde{\Phi}^{''0} = {1 \over 4i} \; [\; (AL-BK - DM + CN) + (A'L'-B'K' - D'M' + C'N')= \tilde{\Phi}^{0}
+ \tilde{\Phi}^{'0} \;  ,
$$
$$
\Psi^{''1} = -{1 \over 4} \;[\;  (AK-BL  + CM-DN ) +  (A'K'-B'L'  + C'M'-D'N' )\; ]= \Psi^{1} + \Psi^{'1} \; ,
$$
$$
\tilde{\Psi}^{''1} =- {1 \over 4i}  \;[\; (AK-BL   - CM + DN) +  (A'K'-B'L'   - C'M' + D'N')\; ] =
\tilde{\Psi}^{1} + \tilde{\Psi}^{'1}\; ,
$$
$$
\Phi^{''2} = -{i \over 4} \; [\;  (AK+BL + CM+DN )+  (A'K'+B'L' + C'M'+D'N' )\; ]=
\Phi^{'2}+ \Phi^{'2} \;  ,
$$
$$
\tilde{\Phi}^{''2} =- {1 \over 4} \; [\; (AK+BL - CM- DN) +  (A'K'+B'L' - C'M'- D'N')\; ]=
\tilde{\Phi}^{2}+ \tilde{\Phi}^{'2}\;  ,
$$
$$
\Phi^{''3} = {1 \over 4}\; [\;  (BK+AL + DM+CN) +  (B'K'+A'L' + D'M'+C'N')\; ] =
\Phi^{3}+ \Phi^{'3}\; ,
$$
$$
\tilde{\Phi}^{''3} = {1 \over 4i}\;[\;  (BK+AL  - DM- CN) +  (B'K'+A'L'  - D'M'- C'N') \; ]=
\tilde{\Phi}^{3} + \tilde{\Phi}^{'3}\; .
\eqno(23b)
$$

\noindent antisymmetric tensor
$$
\Psi ^{''01} ={i \over 4}[\; (AM-BN+CK-DL) + (A'M'-B'N'+C'K'-D'L') \; ] =\Psi ^{01} +\Psi ^{'01}\; ,
$$
$$
\Psi ^{''02} =-{1 \over 4}[\; (AM+BN+CK+DL) + (A'M'+B'N'+C'K'+D'L')\; ] =\Psi ^{02}+ \Psi ^{'02}\; ,
$$
$$
\Psi ^{''03} =-{i \over 4}[\; (AN+BM+CL+DK) + (A'N'+B'M'+C'L'+D'K') \; ] =
\Psi ^{03}+ \Psi ^{'03}\; ,
$$
$$
\Psi ^{''23}={1 \over 4}[\; (AM-BN-CK+DL)  + (A'M'-B'N'-C'K'+D'L')\; ] = \Psi ^{23}+ \Psi ^{'23}\; ,
$$
$$
\Psi ^{''31}={i \over 4}[\; (AM+BN-CK-DL) + (A'M'+B'N'-C'K'-D'L') \; ] =\Psi ^{31} +\Psi ^{'31}\; ,
$$
$$
\Psi ^{''12}=-{1 \over 4}[\; (AN+BM-CL-DK)  + (A'N'+B'M'-C'L'-D'K')\; ]= \Psi ^{12}+ \Psi ^{'12} \; .
\eqno(23c)
$$

It is easily verified that now we have not any additional constraints on
16 tensor components.
Indeed, remembering on relativistic symmetry considerations
we might assume existence of the relationships
$$
\Psi^{''ab} \Psi''_{b}=  \alpha  \tilde{\Psi}'' \tilde{\Psi}^{''a}
+ \rho   \Psi'' \Psi^{''a} \; , \qquad
 \Psi^{''ab} \tilde{\Psi}
''_{b} = \beta \tilde{\Psi}'' \Psi^{''a} + \sigma \Psi''
\tilde{\Psi}^{''a} \; . \eqno(24a)
$$

\noindent
When restricting to only two different 4-spinors, eqs. (24a) give
$$
4\Psi^{ab} \Psi_{b}=  4 \alpha  \tilde{\Psi} \tilde{\Psi}^{a} + 4\rho   \Psi \Psi^{a} \; ,\qquad
4\Psi^{ab} \tilde{\Psi} _{b} = 4\beta \tilde{\Psi} \Psi^{a} + 4\sigma
\Psi   \tilde{\Psi}^{a} \; ;
$$

\noindent
however here an exact solution is known -- therefore  we have identities
$$
\alpha = -1 \; ,  \qquad \beta = +1 \; , \qquad \rho = \sigma = 0 \; .
\eqno(24b)
$$

\noindent
Thus, we are to verify only the following relations
$$
\Psi^{''ab} \Psi''_{b}= - \tilde{\Psi}'' \tilde{\Psi}^{''a}\; ,\qquad
\Psi^{''ab} \tilde{\Psi} ''_{b} = + \tilde{\Psi}'' \Psi^{''a} \; .
\eqno(24c)
$$

It is the matter of simple calculation to show that identities
(24c) do not hold; therefore,  all 16 tensor components of the
Dirac--K\"{a}hler field are independent;.

 Now, the task is to describe restrictions on 4-spinors relevant to pure boson particles
 with fixed spin (0 or 1) and fixed intrinsic parity:

scalar particle
$$
\underline{S=0,}  \qquad \tilde{\Psi}=0, \qquad  \tilde{\Psi}^{a}=0, \qquad  \Psi^{ab}=0, \qquad
\Psi \neq 0, \qquad  \Psi^{a} \neq 0\; ;
\eqno(27a)
$$

pseudo-scalar particle
$$
\underline{\tilde{S} =0,}  \qquad \Psi=0, \qquad  \Psi^{a}=0, \qquad  \Psi^{ab}=0, \qquad
\tilde{\Psi} \neq 0, \qquad  \tilde{\Psi }^{a} \neq 0\; ;
\eqno(27b)
$$

vector particle
$$
\underline{S=0,}  \qquad \tilde{\Psi}=0, \qquad   \Psi=0, \qquad  \tilde{\Psi}^{a}=0,  \qquad
\Psi^{a}  \neq 0, \qquad  \Psi^{ab} \neq 0\; ;
\eqno(27c)
$$

pseudo-vector particle
$$
\underline{S=0,}  \qquad \tilde{\Psi}=0, \qquad   \Psi=0, \qquad  \Psi^{a}=0,  \qquad
\tilde{\Psi}^{a}  \neq 0, \qquad  \Psi^{ab} \neq 0\; ;
\eqno(27d)
$$

First, let us find constrains separating  {\bf scalar particle  (27a)}.  From
$$
 \tilde{\Psi}''  =  - {1\over 4}\;  [\; BM-AN  - CL + DK +  B'M'-A'N'  - C'L' + D'K' \; ]=0
 \; .
$$
it follows
$$
( BM-AN  +  B'M'-A'N' ) =  +(CL - DK    + C'L' - D'K') \; ;
\eqno(28a)
 $$

\noindent
from
$$
\tilde{\Psi}^{''0} = {1 \over 4i} \; [\; AL-BK - DM + CN  + A'L'-B'K' - D'M' + C'N' \; ]= 0  \;  ,
$$
$$
\tilde{\Psi}^{''3} = {1 \over 4i}\;[\;  BK+AL  - DM- CN  +  B'K'+A'L'  - D'M'- C'N'  \; ]=0
$$

\noindent
we get
$$
AL- DM  + A'L'- D'M'  = 0\; , \qquad  BK  - CN  + B'K'  - C'N' =
0\; ; \eqno(28b)
$$

\noindent
restrictions
$$
\tilde{\Psi}^{''1} =- {1 \over 4i}  \;[\; AK-BL   - CM + DN  +  A'K'-B'L'   - C'M' + D'N' \; ] = 0 \; ,
$$
$$
\tilde{\Psi}^{''2} =- {1 \over 4} \; [\; AK+BL - CM- DN  +  A'K'+B'L' - C'M'- D'N' \; ]= 0
$$
give
$$
AK  - CM  +  A'K'  - C'M'  = 0\; , \qquad  BL   - DN  +B'L'    -
D'N' = 0\; ; \eqno(28b')
$$

\noindent
and
$$
\Psi ^{''01} ={i \over 4}[\; AM-BN+CK-DL + A'M'-B'N'+C'K'-D'L' \; ] = 0\; ,
$$
$$
\Psi ^{''23}={1 \over 4}[\; AM-BN-CK+DL  + A'M'-B'N'-C'K'+D'L'\; ] = 0
$$
give
$$
AM-BN + A'M'-B'N'=0\; , \qquad CK-DL +C'K'-D'L'=0\; ; \eqno(28c)
$$

\noindent
restrictions
$$
\Psi ^{''02} =-{1 \over 4}[\; AM+BN+CK+DL  + A'M'+B'N'+C'K'+D'L' \; ] =0\; ,
$$
$$
\Psi ^{''31}={i \over 4}[\; AM+BN-CK-DL  + A'M'+B'N'-C'K'-D'L'  \; ] =0
$$

\noindent
lead to
$$
AM+BN  + A'M'+B'N'=0 \; , \qquad  CK+DL  +C'K'+D'L'= 0 \; ;
\eqno(28c')
$$

\noindent
from
$$
\Psi ^{''03} =-{i \over 4}[\; AN+BM+CL+DK + A'N'+B'M'+C'L'+D'K' \; ] =0
\; ,
$$
$$
\Psi ^{''12}=-{1 \over 4}[\; AN+BM-CL-DK  + A'N'+B'M'-C'L'-D'K'\; ]=0
$$

\noindent it follows
$$
AN+BM+ A'N'+B'M'= 0 \; , \qquad  CL+DK +C'L'+D'K'= 0 \; .
\eqno(28c'')
$$

Thus we have found 11 additional constraints on 16 variables of  four 4-spinors, separating pseudo-scalar particle.

Similarly, for case of   {\bf pseudo-scalar particle  (27b)} we have:
$$
( BM-AN  +  B'M'-A'N' ) =  -(CL - DK    + C'L' - D'K') \; ;
 $$
$$
AL  + DM +  A'L'  + D'M'-C'N'  = 0 \; , \qquad  BK  +CN  +B'K'  +
C'N'  =0 \; ;
 $$
$$
AK + CM  +  A'K' + C'M' = 0 \; , \qquad  BL+DN   +B'L'  +D'N'=0 \;
; \eqno(29a)
$$
$$
AM-BN  +  A'M'-B'N' =0 \; , \qquad CK-DL  +C'K'-D'L'= 0 \; ;
$$
$$
AM+BN  +  A'M'+B'N' = 0 \; , \qquad CK+DL  +C'K'+D'L'  =0 \; ;
$$
$$
AN+BM++  A'N'+B'M''= 0\; , \qquad  CL+DK  +C'L'+D'K'= 0 \; .
\eqno(29b)
$$

\noindent
Again, we have found 11 additional constraints on 16 variables of  four 4-spinors.

In similar manner let us consider the case of the   {\bf vector
particle  (27c)}:
$$
 \Psi'' =-  {1\over 4i} \; [ \; (BM-AN  + CL - DK) + (B'M'-A'N'  + C'L' - D'K')\; ] = 0\; ,\qquad\qquad
$$
$$
 \tilde{\Psi}''  =  - {1\over 4}\;  [\; (BM-AN  - CL + DK) +  (B'M'-A'N'  - C'L' + D'K')\; ]=0
\qquad \Longrightarrow
$$
$$
 BM-AN  +  B'M'-A'N'  = 0 \; , \qquad
CL - DK    + C'L' - D'K'=0 \; ;
\eqno(30a)
$$
$$
\tilde{\Phi}^{''0} = {1 \over 4i} \; [\; (AL-BK - DM + CN) + (A'L'-B'K' - D'M' + C'N')= 0\;, \qquad \qquad
\;  ,
$$
$$
\tilde{\Phi}^{''3} = {1 \over 4i}\;[\;  (BK+AL  - DM- CN) +  (B'K'+A'L'  - D'M'- C'N') \; ]=
\tilde{\Phi}^{3} + \tilde{\Phi}^{'3} \qquad \Longrightarrow
$$
$$
AL - DM + A'L' - D'M' = 0 \; , \qquad BK  - CN  +B'K'  - C'N' =
0\; ; \eqno(30b)
$$
$$
\tilde{\Psi}^{''1} =- {1 \over 4i}  \;[\; (AK-BL   - CM + DN) +  (A'K'-B'L'   - C'M' + D'N')\; ] =0 \; ,
\qquad \qquad
$$
$$
\tilde{\Phi}^{''2} =- {1 \over 4} \; [\; (AK+BL - CM- DN) +  (A'K'+B'L' - C'M'- D'N')\; ]=0
\qquad \Longrightarrow
$$
$$
AK  - CM  +  A'K'  - C'M'   = 0 \; , \qquad BL    - DN    +B'L' -
D'N' =0 \;. \eqno(30c)
$$

Eqs. (30a,b,c) provide us with  6 additional constraints separating
the vector particle.

For the case of  {\bf pseudo-vector particle} we have:
$$
 BM-AN  +  B'M'-A'N'  = 0\; , \qquad
 CL - DK    + C'L' - D'K'=0 \; ;
$$
$$
AL+ DM + A'L'  + D'M' = 0 \; , \qquad BK  +CN  + B'K'   + C'N' = 0
\; ;
$$
$$
 AK  + CM   +  A'K'  + C'M'  = 0 \; , \qquad
   BL  +DN    +B'L'  +D'N' = 0 \; .
\eqno(31)
$$
Again we have obtained 6 constraints on 16 components of four  4-spinors.

Thus, restrictions on four 4-spinors corresponding to separation
of different simple boson with spin 0 or 1 and various intrinsic
parities, are constructed in explicit form.

\section{Conclusions}

Specially note that the use  of four 4-spinor  fields gives possibility to construct the Dirac--K\"{a}hler
tensor set of  16 independent components. However, the formulas relating the Dirac-K\"{a}hler boson to
four fermion fields are completely different from those  previously used in the literature.


\begin{thebibliography}{xxx}

\bibitem{1}
V.I. Strazhev, I.A. Satikov, V.A. Tsionenko.
The Dirac--K\"{a}hler equatin: clfssical field. Minsk: BSU, 2007.

\bibitem{2}
  V.M. Red'kov. Tetrad formalism, spherical symmetry and Schr\"{o}dinger basis.
Publishing House "Belarusian Science", Minsk, 2011.



\end{thebibliography}
\end{document}